\documentclass[preprintnumbers,twocolumn,amsmath,amssymb,floatfix,9pt,prd,onecolumn,
superscriptaddress,nofootinbib]{revtex4}
\usepackage{latexsym}
\usepackage{epsfig}
\usepackage{amssymb}

\begin{document}

\title{A new deterministic model of   strange stars}

\author{Farook Rahaman}
\email{rahaman@iucaa.ernet.in} \affiliation{Department of
Mathematics, Jadavpur University, Kolkata 700032, West Bengal,
India}

\author{ Koushik Chakraborty}
\email{kchakraborty28@yahoo.com } \affiliation{Department of
Physics, Government Training College, Hooghly - 712103, West
Bengal, India}

\author{P.K.F. Kuhfittig}
\email{kuhfitti@msoe.edu} \affiliation{Department of Mathematics,
Milwaukee School of Engineering, Milwaukee, Wisconsin 53202-3109,
USA}

\author{ G. C. Shit}
\email{gopal_iitkgp@yahoo.co.in } \affiliation{Department of
Mathematics, Jadavpur University, Kolkata - 700032, West Bengal,
 India}

\author{Mosiur Rahman}
\email{mosiurju@gmail.com}\affiliation{Department of Mathematics,
Meghnad Saha Institute of Technology, Kolkata-700150, India}
\date{\today}

\begin{abstract}\noindent
{\bf Abstract:} The observed evidence for the existence of strange
stars and the concomitant observed masses and radii are used to
derive an interpolation formula for the mass as a function of the
radial coordinate.  The resulting general mass function becomes
an effective model for a strange star.  The analysis is based on
the MIT bag model and yields the energy density, as well as the
radial and transverse pressures. Using the interpolation function
for the mass, it is shown that a mass-radius relation due to
Buchdahl is satisfied in our model. We find the surface redshift
($Z$) corresponding to the compactness of the stars.  Finally,
from our results,  we   predict some characteristics of a strange
star of  radius 9.9 km.   \\

\phantom{a}


\end{abstract}

\maketitle

\section{Introduction}\noindent
The constitution of the interior of neutron stars is still considered
an open question by the scientific community.  Not only are neutron
stars highly dense compact astrophysical objects, the extreme
conditions near the center have led to the hypothesis that the core
consists entirely of \emph{quark matter}, in which the constituent
quarks comprising the neutrons become deconfined.  Moreover, based
on the MIT bag model, it was first argued explicitly by Witten
\cite{witten} that the most stable state of matter is \emph{strange
quark matter}, which is a mixture containing roughly the same
number of up, down, and strange quarks.

Quark matter is of interest for a number of reasons.  For example,
that compact stars could result in a topology change, that is, in
the formation of wormholes, had already been suggested in Ref.
\cite{aD08}.  It is shown in Ref. \cite{pK13} that the topology
change inside a neutron star requires a quark-matter core of a
certain minimal radius.  The survival of quark nuggets from the
Early Universe is also a possibility.  Using the MIT bag model,
it is shown in Ref. \cite{fR12} that quark matter may be a
suitable candidate for dark matter. In fact, a strange star may
be regarded as a huge strangelet.

In QCD, the interaction between  quarks becomes weak for a large
exchange of momentum. So, for a  sufficiently large temperature or
density,  or both,  interaction between constituent quarks becomes
very weak and,  consequently, they become deconfined. In heavy ion
collider experiments deconfinement of quarks may be brought about
at high temperatures ($\sim$ 180 MeV or above). But neutron stars
are cold ($\sim$ few KeV). Extremely large chemical potential at
the core of the neutron star plays the central role for
deconfinement of the quarks. At densities of nearly twice nuclear
density,  hyperons appear in neutron star matter and this state is
known as the hadronic phase (HP),  whereas a  deconfined quark
matter phase is obtained as a phase transition from the hadronic
phase at densities much higher than the nuclear density.

If the conversion of neutron-star matter to quark matter is not
confined to the core, then the result is a \emph{quark star}.
Under certain conditions, it is theoretically possible for some
up and down quarks to be transformed into strange quarks. Since
the strange matter is the true ground state of matter, nothing
can stop the conversion of the entire quark star into starnge
matter once the core gets converted into strange matter. Thus a
neutron star gets converted into a strange star.  In this paper,
we propose a new deterministic model for strange
stars based on the MIT bag model.\\
 We organize our paper as follows:\\
  In Sec II, we have provided the basic equations. In Sec. III,
   we have obtained the solutions of physical parameters. In section
   IV, we have studied
mass-radius relation \& surface redshift  of the stranger stars.
The article is concluded with a short discussion.

\section{Basic Equations}\noindent
To describe the spacetime of the interior of the strange star,  we assume
the metric to be
\begin{equation}
ds^{2}=-e^{\nu(r)}dt^2+e^{\lambda} dr^2+r^2(d\theta^2+\sin^2\theta
\,d\phi^2) \label{eq:kbm}
\end{equation}
and then recall that the most general energy momentum tensor compatible
with spherically symmetry is
\begin{equation}
T_\nu^\mu=  ( \rho + p_r)u^{\mu}u_{\nu} - p_r g^{\mu}_{\nu}+
            (p_t -p_r )\eta^{\mu}\eta_{\nu} \label{eq:emten}
\end{equation}
with
\[
 u^{\mu}u_{\mu} = - \eta^{\mu}\eta_{\mu} = 1.
 \]

The Einstein field equations are listed next.
\begin{equation}e^{-\lambda}
\left[\frac{\lambda^\prime}{r} - \frac{1}{r^2}
\right]+\frac{1}{r^2}= 8\pi \rho , \label{eq:lam}
\end{equation}
\begin{equation}e^{-\lambda}
\left[\frac{1}{r^2}+\frac{\nu^\prime}{r}\right]-\frac{1}{r^2}=
8\pi p_r, \label{eq:nu}
\end{equation}
\begin{equation}\frac{1}{2} e^{-\lambda}
\left[\frac{1}{2}(\nu^\prime)^2+ \nu^{\prime\prime}
-\frac{1}{2}\lambda^\prime\nu^\prime + \frac{1}{r}({\nu^\prime-
\lambda^\prime})\right] =8\pi p_t.  \label{eq:tan}
\end{equation}
Our analysis begins with a set of astrophysical objects considered
to be candidates for strange stars.  The masses and radii of these
compact objects are listed in Table 1.  The interpolation technique has been used to estimate the cubic polynomial that yield the following expression for the mass
as a function of the radial coordinate $r$:

\begin{table*}
\centering \caption{The values of the mass and radius for  various
strange stars  \cite{DR,note}  }

\begin{tabular}{|c|c|c|l|} \hline
Strange Stars & Radius ( in km)   & Mass ($M_{\odot}$) & The  mass in km   (1 $M_{\odot} = 1.475$ km) \\
\hline PSR J1614-2230 & 10.3
 & 1.97 $\pm$ 0.04& ~~~~~~~~~~~~~~~~~~2.9057 $\pm$ 0.059\\
\hline Vela X - 1  & 9.99 & 1.77 $\pm$ 0.08& ~~~~~~~~~~~~~~~~~~2.6107 $\pm$ 0.118\\
\hline  PSR J1903+327 & 9.82
 & 1.667 $\pm$ 0.021 & ~~~~~~~~~~~~~~~~~~2.4588 $\pm$ 0.03\\ \hline Cen X - 3  &   9.51&1.49 $\pm$ 0.08& ~~~~~~~~~~~~~~~~~~2.1977 $\pm$ 0.118 \\
 \hline SMC X - 1 &   9.13&1.29 $\pm$ 0.05& ~~~~~~~~~~~~~~~~~~1.9027 $\pm$ 0.073 \\
  \hline\end{tabular}
\end{table*}

 \begin{equation}
m ( r ) = a r^3 - b r^2 + c r - d,
\label{eq:mass}
\end{equation}
where $a = 0.01492 $, $b = 0.3296$, $c = 3.03$, and $d = 9.6453$.
The graph of $m(r)$ is shown in Fig. 1 with residuals corresponding to the
observed data and fitted cubic polynomial. From this figure, we notice that
 the maximum norm of residuals is about $10^{-5}$.   The extended range of
  the mass radius relation is shown in Fig.2.  Observe that the best
fit is obtained in the range around 7 km to 12 km. We will see
later that the accepted range is $r=6.2\,\,\text{km}$ to
$r=12.2\,\,\text{km}$.  It is generally well known that for any
interpolation technique the degree of the polynomial increases
with the increase of data points. However, spline interpolation
resolves this problem. Because the spline interpolation is a
special type of piecewise polynomial so that the interpolation
error can be made small even when using low degree polynomials.
In our problem, a spline interpolation polynomial is constructed
using the given set of observed data for wide range of radius of
the stars. Afterwards we have fitted a cubic polynomial (for mass
radius relation) using the data extracted from spline
interpolation. Therefore, use of additional one or more observed
data will not significantly affect the results presented here.

\begin{figure}
    \centering
        \includegraphics[scale=.7]{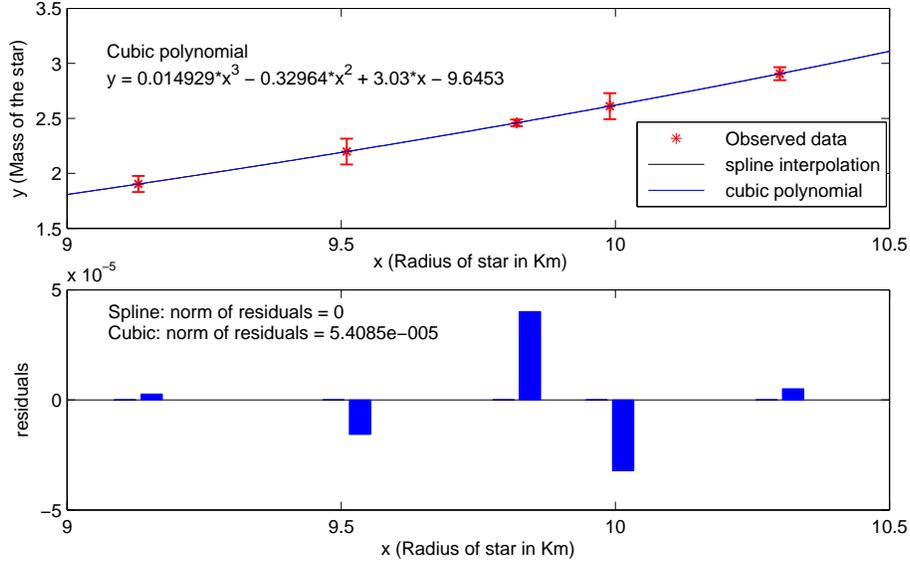}
    \caption{Interpolation curves from the observed data of the strange-star
    candidates with residuals.}
    \label{Fig 1}
\end{figure}

\begin{figure}
    \centering
        \includegraphics[scale=.7]{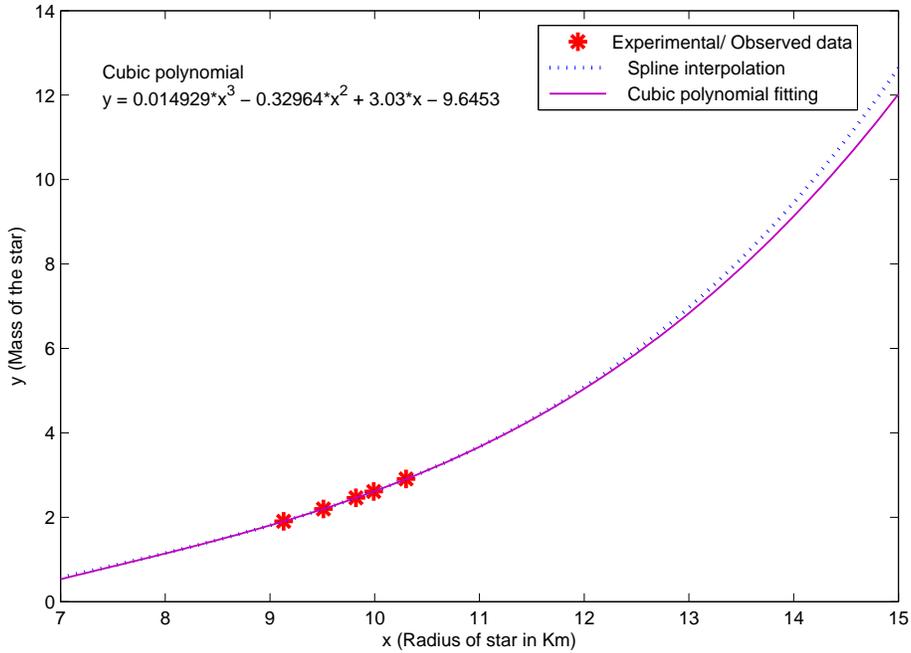}
    \caption{Interpolation curves from the observed values of mass
      and radius of the strange-star candidates with extended ranges.}
    \label{Fig 1}
\end{figure}
Further analysis is going to be based on the MIT bag model.
In this model, the strange matter is assumed to have the
following equation of state \cite{RSRMK}:
\begin{equation}\label{bag}
p_r = \frac{1}{3}(\rho - 4B),
\end{equation}
where $B$, the bag constant, is in units of $\text{MeV/(fm)}^3$.
To obtain a value suitable for our analysis, it is important to
note that for the strange-star candidates, $B$ has been found
to lie in the range 60-80 $\text{MeV/(fm)}^3$ for a
$\beta$ - equilibrium stable strange-matter configuration
\cite{Farhi, Alcock}.  A convenient value for plotting purposes
is $B=0.0001$.  Since this choice corresponds to
83 $\text{MeV/(fm)}^3$, it is close to the accepted values.
Indeed,
\[
   83\frac{\text{MeV}}{(\text{fm})^3}\frac{G}{c^4}
   \times (10\,\text{m})^6=0.0001 (\text{km})^{-2}.
\]
For other possible values of $B$, see Ref. \cite{Kalam}.

\section{Solutions}\noindent
From the metric potential $e^{\lambda}$ in Eq. (\ref{eq:lam}), we get
\begin{equation}\label{grr}
   e^{{-\lambda}}=1-\frac{2m(r)}{r} = 1 - 2ar^2 + 2br - 2c
      + \frac{2d}{r}.
\end{equation}
Next, making use of Eq. (\ref{bag}), we obtain from the Eqs.
(\ref{eq:lam}), (\ref{eq:nu}), and
    (\ref{eq:tan}),
\begin{equation}\label{density}
\rho = \frac{1}{8\pi}\left( 6a - \frac{4b}{r} +
   \frac{2c}{r^2} \right),
\end{equation}

\begin{equation}\label{radial}
  p_r = \frac{1}{24\pi}\left( 6a - \frac{4b}{r} +
  \frac{2c}{r^2} \right) - \frac{4B}{3},
\end{equation}
and


\begin{equation}\label{pt}
p_{t} = \frac{1}{16\pi}\left( 1 - 2ar^2 + 2br - 2c + \frac{2d}{r}
\right)\left( I_1 - I_2 + I_3 + I_4 - I_5 \right),
\end{equation}
  where \[ I_1 = \frac{\left(12ar^3 - 10br^2 - 4cr -
\frac{Br^3}{2\pi} -6d \right)}{2\left(3r^2 - 6ar^4 + 6br^3 -
6cr^2 + 6dr \right)} ,\]

\[ I_2 = \frac{(4ar - 2b + \frac{2d}{r})(12ar^3 - 10br^2 - 4cr
-\frac{Br^3}{2\pi} - 6d)}{2(1 - 2ar^2 + 2br - 2c +
\frac{2d}{r})(3r^2 - 6ar^4 + 6br^3 - 6cr^2 + 6dr)},\]

\[  I_3 = \frac{\frac{12ar^3 - 10br^2 - 4cr -\frac{Br^3}{2\pi} -
6d}{3r^2 - 6ar^4 + 6br^3 - 6cr^2 + 6dr} - \frac{4ar - 2b +
\frac{2d}{r^2}}{1 - 2ar^2 + 2br - 2c + \frac{2d}{r}}}{r} ,\]

\[ I_4 = \frac{36ar^2 - 20br - 4c - \frac{3Br^2}{2\pi}}{3r^2 -
6ar^4 + 6br^3 - 6cr^2 + 6dr}, \]

\[I_5 = \frac{(12ar^3 - 10br^2 - 4cr - \frac{3Br^3}{2\pi} - 6d)(6r
- 24ar^3 + 18br^2 - 12cr + 6d)}{(3r^2 - 6ar^4 + 6br^3 - 6cr^2 +
6dr)^2}. \]


\begin{figure*}
\begin{tabular}{rl}
\includegraphics[width= 5cm]{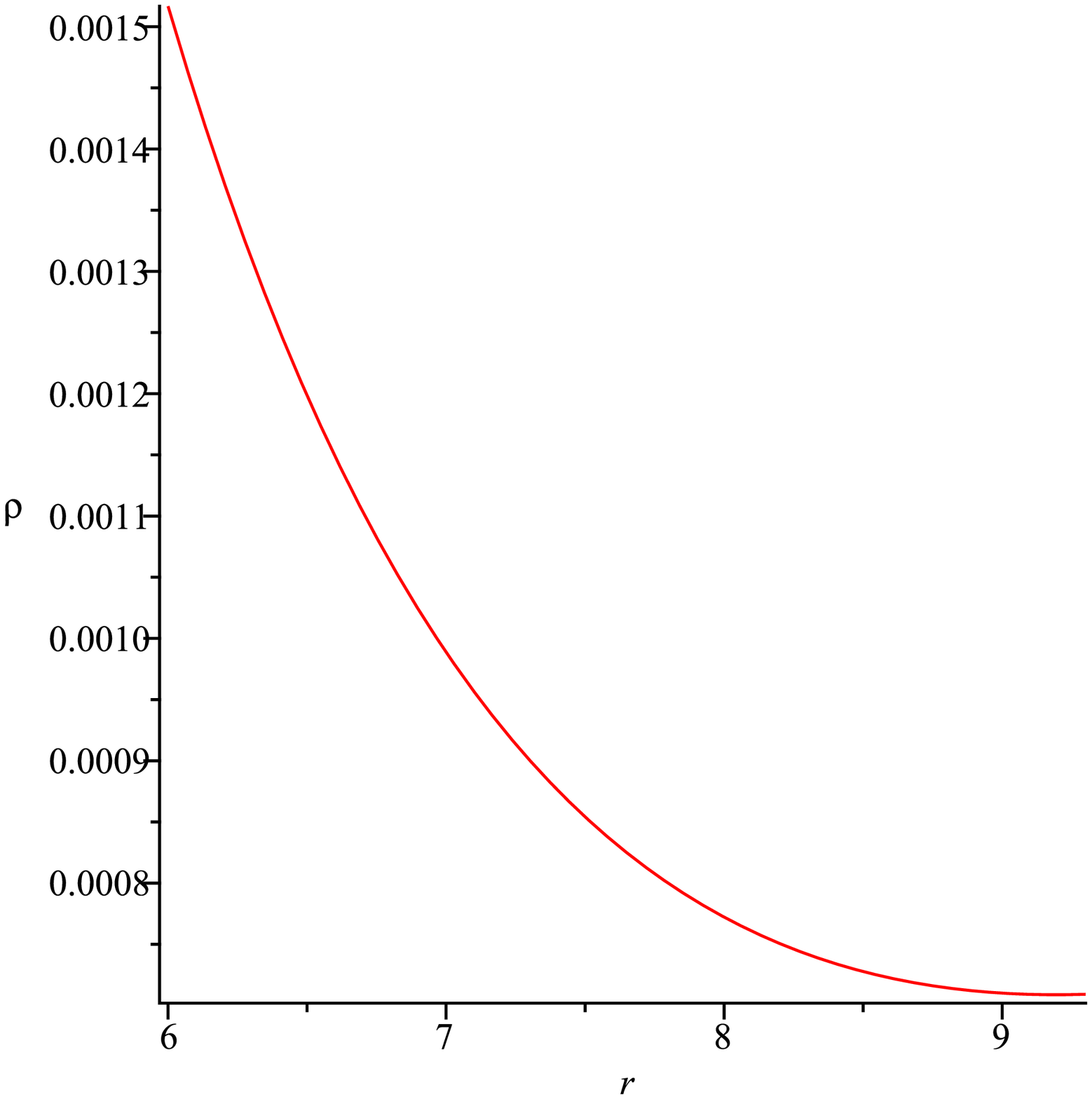}&
\includegraphics[width= 5cm]{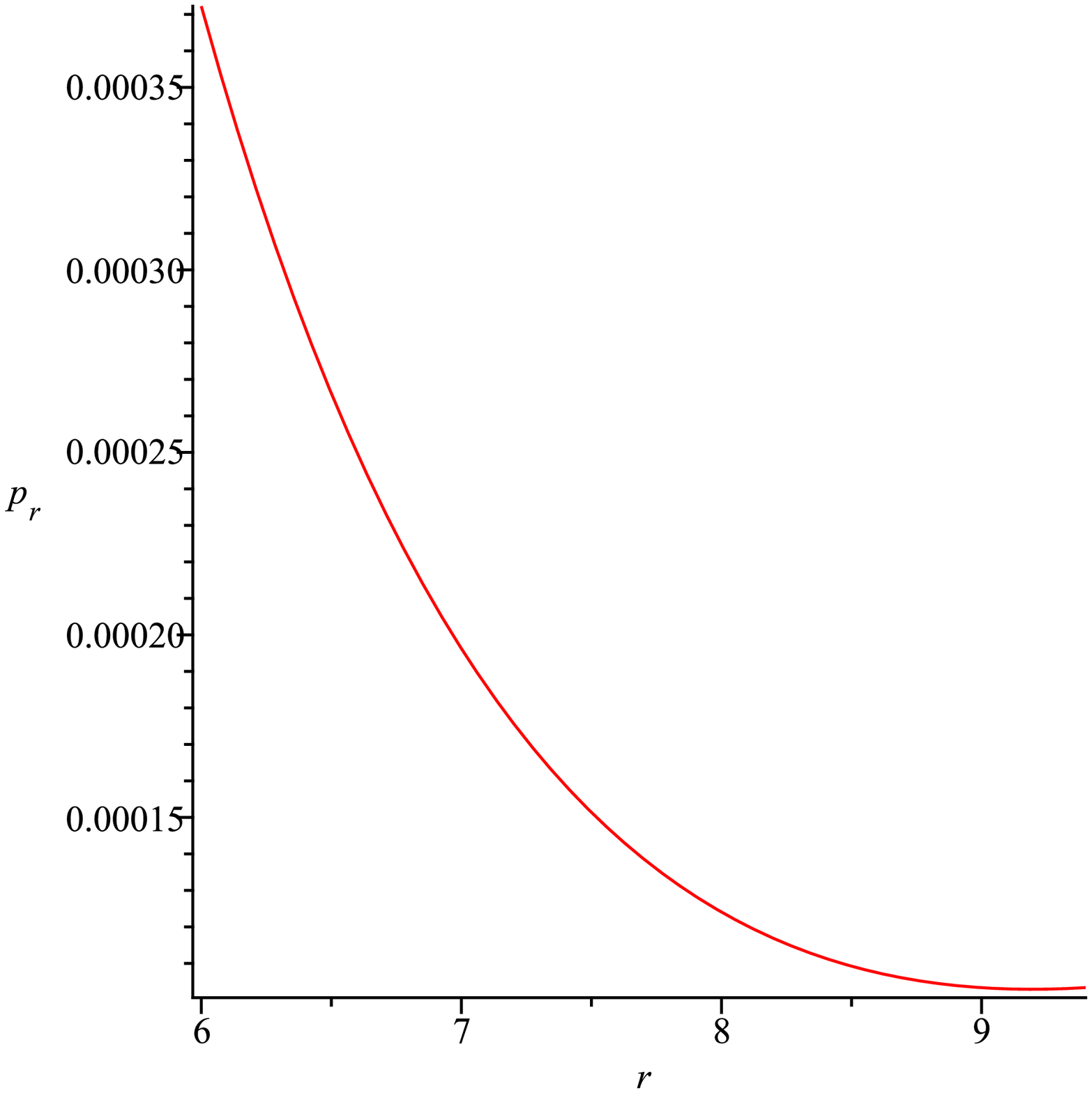}
 \includegraphics[width= 5cm]{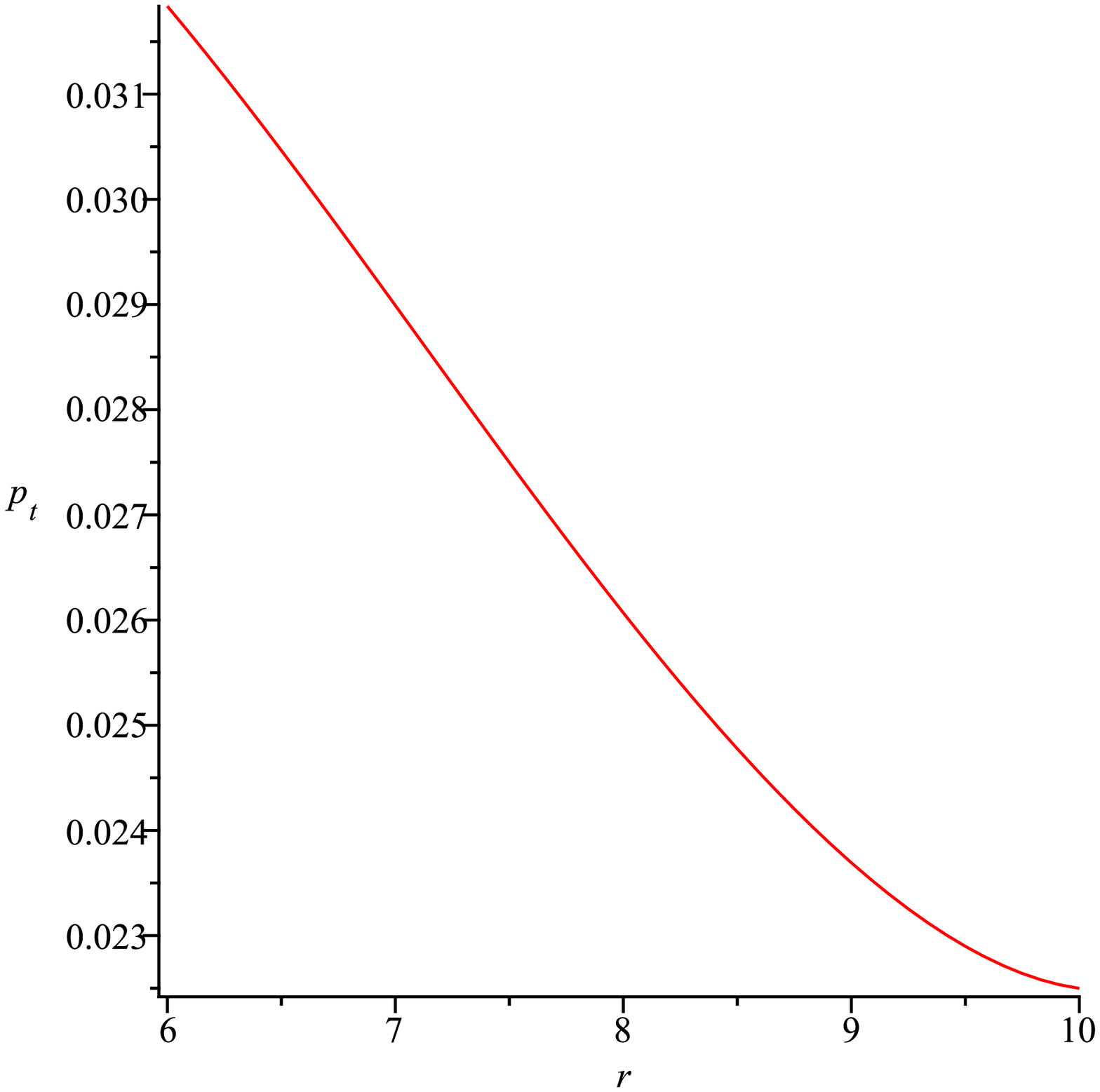} \\
\end{tabular}
\caption{ (\textit{Left}) The plot of $\rho$ as a function of $r$
(in km)  (\textit{Middle}) Plot of $p_r$  vs. $r$
  (in km) with $B = 10^{-4}$.
   (\textit{Right}) Plot of $p_t$ vs. r (in km) with
     $B = 10^{-4}$.}
\end{figure*}

The plots  of Eqs. (\ref{density})- (\ref{pt}) are shown in Figs.
2.

\begin{table*}
\centering \caption{Verification of Buchdahl's condition for
various radii}
\begin{tabular}{|c|c|l|} \hline
Radius(km)&$\frac{2M}{R}$&Comments\\
\hline 6.1 & -0.013 & negative \\  \hline 6.2 & 0.0086 &$ \leq
0.89$
\\  \hline
\hline
7 & 0.15196 & $\leq 0.89$\\ \hline 8 & 0.28483 &$ \leq 0.89$\\
\hline 9 & 0.4008 & $\leq 0.89$\\ \hline 10 & 0.5229 & $\leq
0.89$\\ \hline 11 & 0.6657& $\leq 0.89$\\ \hline  \hline 12.2 &
0.8779  &$ \leq 0.89$  \\ \hline 12.3 & 0.8999  &$ \geq 0.89$\\
\hline
 \end{tabular}
\end{table*}

\section{Mass-Radius Relation and Surface redshift }\noindent
Our final task is to study the maximum allowed mass-to-radius
ratio in our model.  Buchdahl \cite{Buchdahl} showed that for a
perfect-fluid sphere, twice the ratio of the maximum allowed
mass to the radius is 8/9, i.e., $2M/R\le 8/9$.  Moreover,
if the trace of the energy-momentum tensor is postulated
to be nonnegative, then the ratio of the total mass to the
coordinate radius is $\le 5/18$.  In other words, $M/R$ is
strictly less than 4/9.  Now, while we would expect the
quantity $1-2M/R$ to be nonnegative, Buchdahl's condition
actually does not allow the value to be less than 1/9.

To see how these conditions apply to our model, we return to Eq.
(\ref{eq:mass}) and recall that the best fit occurs between the
radii 6.2 km and 12.2 km.  In this range, the condition
$1-2M/R>1/9$ is met, as required by Buchdahl's condition.  As can
be seen from the Table II that for different radii this condition
is obeyed in our model of  strange stars.
\\
\begin{table*}
\centering \caption{Redshift for   various radii}
\begin{tabular}{|c|l|} \hline
Radius(km)&$Z$ \\
7 & 0.0859  \\ \hline 8 & 0.1824\\
\hline 9 & 0.29189\\ \hline 10 & 0.4478 \\
\hline 11 & 0.7296 \\ \hline 11.5& 0.9924 \\ \hline
\end{tabular}
\end{table*}

\pagebreak

The compactness of the star is obtained as
\begin{equation}\label{grr}
     u= \frac{ m(r)}{r} =    ar^2-   br + c
      -\frac{ d}{r}.
\end{equation}
The nature of the  compactness of the star is shown   in Fig.
3(left). The surface redshift ($Z$) corresponding to the above
compactness ($u$) is given by
\begin{equation}
\label{eq36} 1+Z= \left[ 1-(2 u )\right]^{-\frac{1}{2}} ,
\end{equation}
where
\begin{equation}
\label{eq37} Z= \frac{1}{\sqrt{2ar^2- 2br +2c
      -\frac{2d}{r}}}-1
\end{equation}
The redshift Z can be measured from the X-ray spectrum. This Z
actually gives the compactness  of the star. High observed
redshifts (0.35-0.45) are consistent  with strange stars which
have mass-radius ratios  higher than neutron stars. Thus, it is
easy to find the maximum surface redshift for the anisotropic
strange stars of different radii  from equation (14). We calculate
the maximum surface redshift for  different strange stars with
different radii,   which is shown in Table III. The nature of the
surface redshift  of the star is shown   the Fig. 3 (right). Note that since, at the surface,  radial pressure is zero i.e. $p_r(r=R)$, therefore, this equation gives a relation between  parameters a,b,c,d and   the parameter B. Therefore, properties of strange matter (comprising Bag constant, B ) enter in the redshift value.

\begin{figure*}
\begin{tabular}{rl}
\includegraphics[width= 5cm]{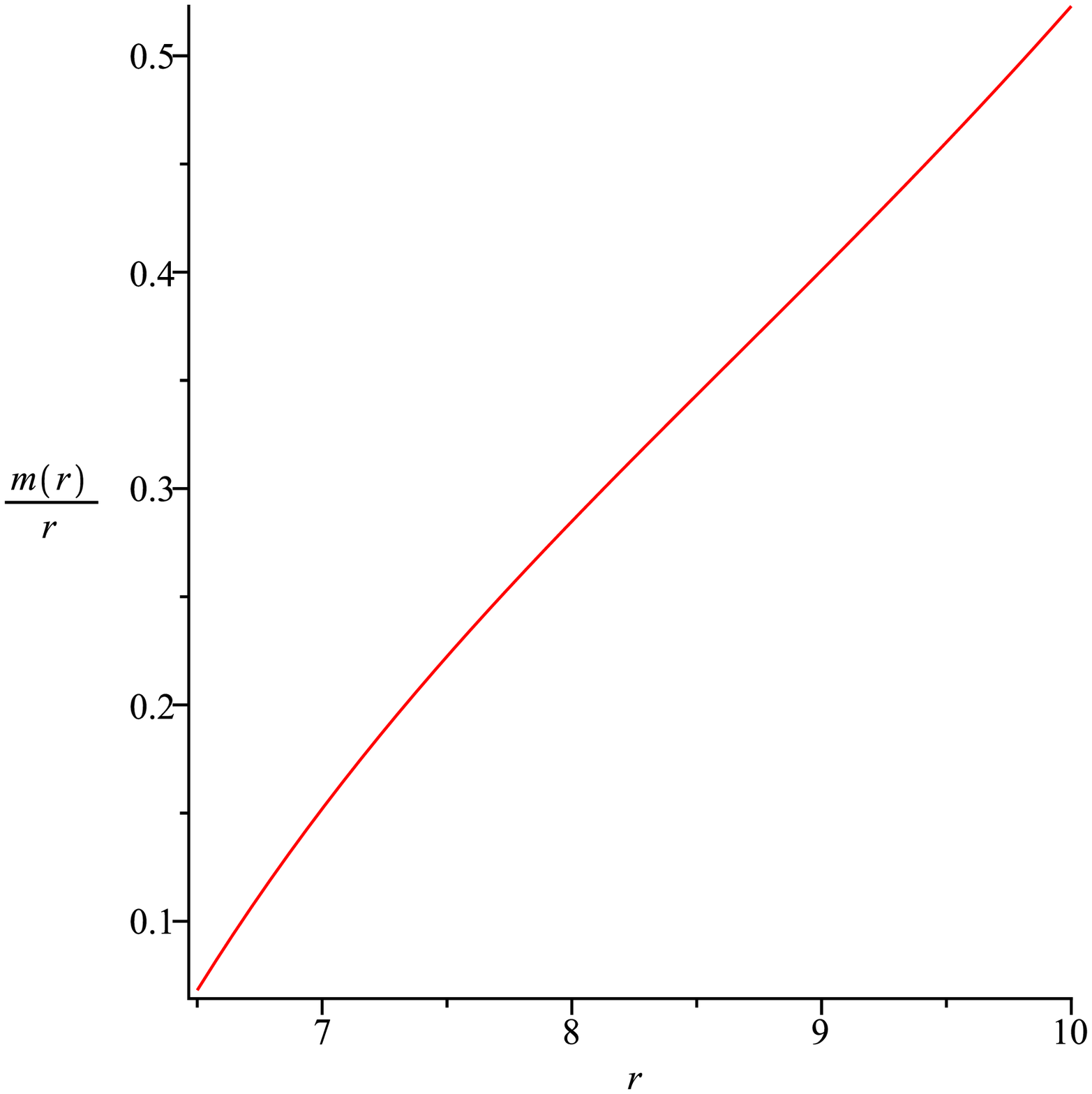}&
  \includegraphics[width= 5cm]{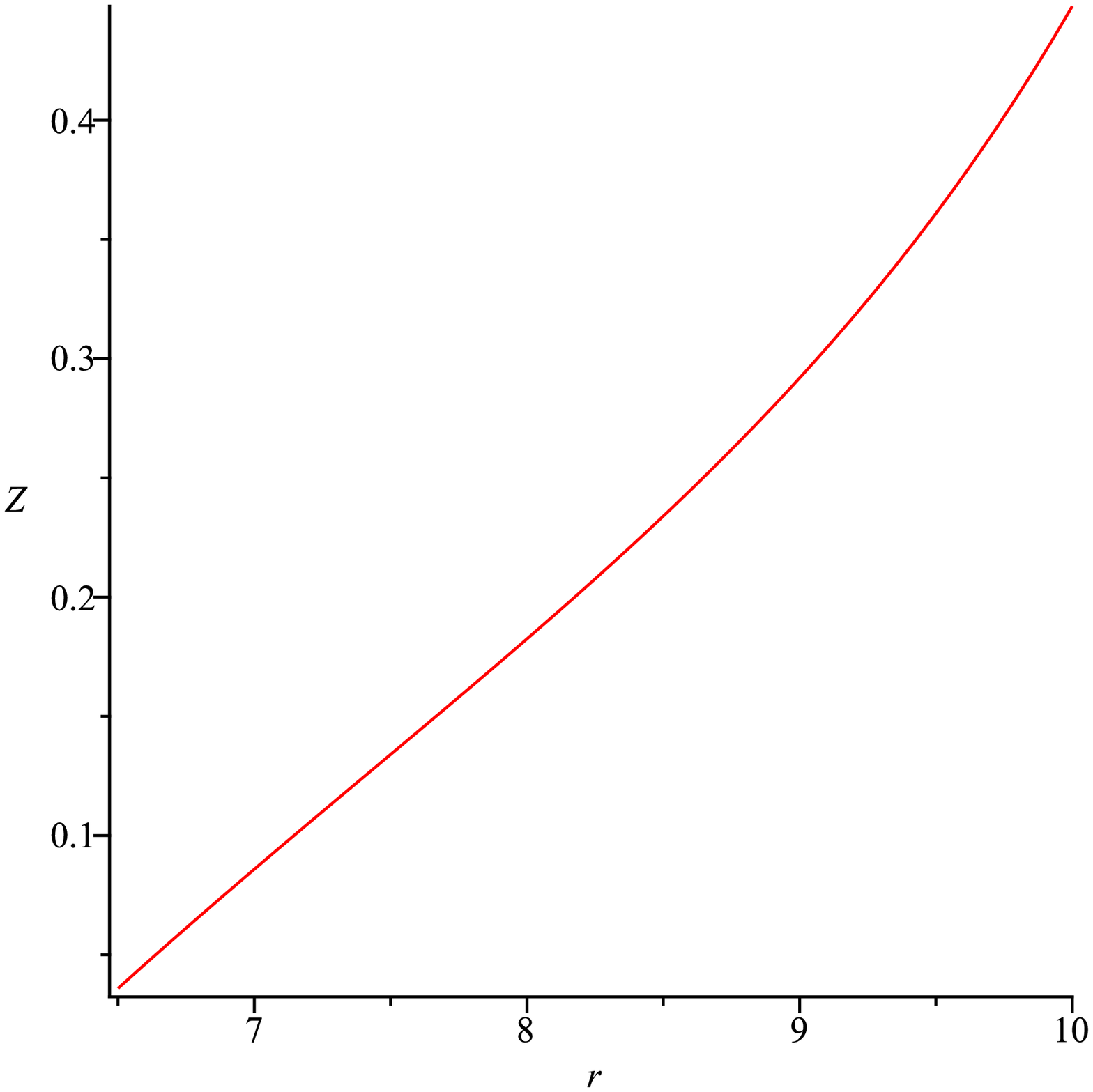} \\
\end{tabular}
\caption{ (\textit{Left}) The variation of $ \frac{m(r)}{r}$ with
respect to  $r$ ( in km ).
   (\textit{Right}) The variation of  the redshift  function with respect to  $r$ ( in km ).}
\end{figure*}

\section{Conclusion}\noindent
The observed evidence for the existence of strange stars has also
led to observed masses and radii.  These observations are used in
this paper to obtain an interpolation function $m(r)$, which
proved to be an effective model for strange stars.  The
subsequent analysis is based on the MIT bag model and yields both
the radial pressure $p_r$ and the transverse pressure $p_t$, as
well as the energy density of the strange star.  Subsequently,
$m(r)$ was used to show that a mass-radius relation due to
Buchdahl is satisfied in our model of a strange star. From our
results we can predict some characteristics of a  strange star of
radius, say 9.9 km. One can see that  we have set $c=G=1$ in our
calculations.
 Now, if one substitutes   $G$  and $c$ into relevant equations, the
  value of the surface density of the predicted  strange star of radius 9.9 km
   turns out to be $ \rho_s =
  0.23~\times ~ 10^{15}$ ~ gm ~cm$^{-3}$. Note that our model  cannot
  predict the central density since we cannot   get the values of the
  physical parameters less than 6.2 km radius.
 Also, we can predict the mass of   strange star  of radius 9.9 km   as $
1.71 M_\odot $.  The compactness of the star will be 0.2548 and
corresponding redshift is Z=0.4285. This high
  redshift   is  convenient  for explaining strange
stars. We can definitely state that  our predicted  strange star
is more compact than neutron stars. In 2008, Cackett et al.
\cite{Cackett} reported  that redshift of a  strange star in the
low-mass X-ray binary 4U 1820-30 is $Z = 0.43$.  This supports
our prediction on strange stars   in the low-mass X-ray binary 4U
1820-30. Recently, X-ray binaries XTE J1739-285 were suggested as
strange stars \cite{MT}. We hope our method can be used to
determine different characteristics of these strange stars.

\section*{Acknowledgements}\noindent FR and KC wish to thank the
authorities of the Inter-University Centre for Astronomy and
 Astrophysics, Pune, India, for providing them Visiting Associateship
under which a part of this work was carried out. FR is also
thankful to UGC, DST for providing financial support. KC is
thankful to University Grants Commission to provide financial
assistance under MRP with which this work was carried out We are
very grateful to an anonymous referee for his insightful comments
and constructive suggestions that  have  led to significant
improvements, particularly on the interpretational aspects.


\begin{thebibliography}{99}
\bibitem{note}  The extreme condition of pressure and density at the
 center of the Neutron Stars may lead to phase transition of the neutrons to hyperons
 \cite{AWG}, bosons, quark matter or strange quark matter
 \cite{EW}.  However, the maximum mass predicted from
  the consideration of the existence of hyperons or bosons, is much smaller than the
  recently discovered compact objects like J1614-2230 \cite{PD} or J0348 +0432 \cite{JA}.
   These observations practically rules out the possibility of existence of hyperons,
   bosons or pure neutrons inside the 'so called' Neutron Stars. The hadrons inside these
    stars must get converted to quark or strange quark matter.
So, though these compact star candidates are called Neutron
Stars, the theoretical predictions of QCD confirm the
transformation of hadrons to quark or strange quark matter.
Hence, from theoretical consideration we are safe to consider
these compact objects as Strange Stars.
\end{thebibliography}
\end{document}